\documentclass[pre,aps,showpacs,twocolumn,floatfix,superscriptaddress]{revtex4}
\usepackage{epsfig}
\usepackage{color}
\usepackage{amsmath,amssymb}

\newcommand{\be}{\begin{equation}}
\newcommand{\ee}{\end{equation}}
\newcommand{\bea}{\begin{eqnarray}}
\newcommand{\eea}{\end{eqnarray}}
\newcommand{\ba}{\begin{array}}
\newcommand{\ea}{\end{array}}

\begin{document}
\title{Translocation time of periodically forced polymer chains}

\author{Alessandro Fiasconaro}
\email{afiascon@unizar.es}
\affiliation{Departamento de F\'{\i}sica de la Materia Condensada,  Universidad de Zaragoza, 50009 Zaragoza,  Spain}
\affiliation{Centro Universitario de la Defensa de Zaragoza, Ctra. de Huesca s/n, E-50090 Zaragoza, Spain}
\affiliation{Dipartimento di Fisica e Tecnologie Relative,  GIP, Universit\`a di Palermo,  Viale delle Scienze, I-90128 Palermo, Italy}

\author{Juan Jos\'e Mazo}
\affiliation{Departamento de F\'{\i}sica de la Materia Condensada, Universidad de Zaragoza, 50009 Zaragoza, Spain} \affiliation{Instituto de Ciencia de Materiales de Arag\'on,  CSIC-Universidad de Zaragoza, 50009 Zaragoza, Spain}

\author{Fernando Falo}
\affiliation{Departamento de F\'{\i}sica de la  Materia Condensada, Universidad de Zaragoza, 50009 Zaragoza, Spain}
\affiliation{Instituto de Biocomputaci\'on y F\'{\i}sica de Sistemas  Complejos, Universidad de Zaragoza, Zaragoza, Spain}

\date{\today}
\begin{abstract}
We show the presence of both a minimum and clear oscillations in the
frequency dependence of the translocation time of a polymer described
as a unidimensional Rouse chain driven by a spatially localized
oscillating linear potential. The observed oscillations of the mean
translocation time arise from the synchronization between the very
mean translocation time and the period of the external force. We have
checked the robustness of the frequency value for the minimum
translocation time by changing the damping parameter, finding a very
simple relationship between this frequency and the correspondent
translocation time. The translocation time as a function of the
polymer length has been also evaluated, finding a precise $L^2$
scaling. Furthermore, the role played by the thermal fluctuations
described as a Gaussian uncorrelated noise has been also investigated,
and the analogies with the resonant activation phenomenon are
commented.
\end{abstract}


\maketitle

\section{Introduction}
The transport features of molecules and polymers attract nowadays more
and more interest. In particular, the translocation features of DNA
polymers through pore cells have been studied with great attention
\cite{KasPNAS96} and different models have been introduced to describe
such process. In between them we can mention for instance the single
barrier potential \cite{Pizz} and the flashing ratchet models
\cite{linke}.  Despite all these efforts, the basic understanding of
the translocation process of polymers is still missing.  In many
cases, the translocation phenomena involve also molecular motors
\cite{Bust,gut} whose complex action has been only recently addressed
with high attention.  In addition, nanotechnological machines can
learn from the biological processes and thus to work in a simpler and
more efficient way \cite{mickler}.  The goal of this article is to
study a cyclic time-dependent model that can depict a simple
nanotechnological machine which is able to drive with a sinusoidal
force a polymer chain in one direction. This simple object
constitutes, in our purpose, a basic model that will be further
developed into a more realistic machine able to emulate the basic
behavior of a biological motor. We find a non trivial behavior of the
polymer translocation time as a function of the driving frequency
$\nu$ with the occurrence of a series of oscillations. These
oscillations are not related to the vibrational internal modes of the
chain but to the driving frequency. The first minimum occurs at mean
times $\tau_m \lesssim T_m$, where $T_m=1/\nu_m$ is the period of the
driving at this minimum. In this sense, the translocation time
dependence presents some similarities with the resonant activation
phenomena \cite{RA}. The polymer scaling properties have been checked
and verified as a function of both the number of polymer beads, and
the damping. The article is structured in the following way: next
section presents the model, then we report the results in Sec.III. In
Subsec.~IIIc the dependence of the stall force of the system is also
studied, finding a strong nonmonotonic behavior with the frequency
even in the presence of relatively high damping. We finish with a
conclusions section.

\section{The model}

\begin{figure}[htbp]
\centering
\includegraphics[width=8.5cm]{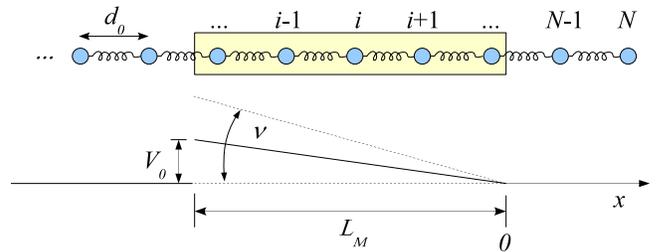}
\caption{(Color online) Scheme of the oscillating pushing force acting
  on a polymer chain formed by N monomers. The force drives only a
  small fraction of the whole chain.} \label{schema}
\end{figure}
The polymer is modeled as a unidimensional chain of $N$
dimensionless monomers connected by harmonic springs \cite{Rouse},
whose total potential energy is
 \be
 V_{\rm har}=\frac{k}{2}\sum_{i=1}^{N-1} (x_{i+1}-x_i-d_0)^2.
 \label{v-har}
 \ee
\noindent Here $k$ is the chain elastic constant, $x_i$ the position
of the $i$-th monomer and $d_0$ the equilibrium distance between
adjacent monomers.

The translocation is helped by the presence of a linear potential (see
Fig.~\ref{schema}) oscillating with frequency $\nu$ which acts in a
space region of length $L_M$. In particular:
\be
V_{\rm sp}(x,t) = \left\{
   \begin{array} {ll}
      - F_0 x \left[ 1 - \cos(2\pi\nu t + \phi) \right],  & x \in [-L_M, 0]  \vspace{0.2cm}\\
      0, & {\rm otherwise}
   \end{array} \right.
   \label{v-sp}
 \ee
\noindent The dynamics of the $i^{\rm th}$ monomer of the chain is then
described by the following underdamped Langevin equation:
 \be
  \ddot{x}_i + \gamma \dot{x}_i =
  -\frac{\partial{V_{har}}}{\partial{x_i}}-\frac{\partial{V_{sp}}}{\partial{x_i}} + \xi_i(t) \label{Lang}.
 \nonumber
 \ee
\noindent The polymer is subject to a viscosity parameter $\gamma$,
being neglected other more complex hydrodynamical effects. Thermal
fluctuations are described as Gaussian uncorrelated noise $\xi_i(t)$
which satisfies the usual statistical properties
$\langle\xi_i(t)\rangle=0$ and $\langle\xi_i(t)\xi_j(t+\tau)\rangle =
2\gamma D \delta_{i j}\delta(\tau)$ with $(i,j=1...N)$. $F_0$ and
$\nu$ represents respectively the amplitude and the frequency of the
forcing field, and $\phi$ its initial phase.

The model here described represents a simple example of a driven polymer translocation, which may reveal as proper mechanism in nanotechnological devices and as a reference example in the area of the translocation driving machines. With this we mean a system which is able to push a polymer chain from one point to another, in analogy with a molecular motor which push a long molecule in the biological translocation process.

\section{Results}\label{Results}
In this manuscript we will focus on studying the dependence of the mean first passage time, as defined below, and of the stall force of the polymer with the frequency of the external driving. Simulations were done at different values of damping and noise (usually $\gamma=1.0$ and $D=0.01$) for $F_0=0.1$, $k=1$ and $L_M=5.5$. We have studied chains with $N$=12,18,24,36 and 120 monomers. The computed mean values are the result of averaging over $10^4$ realizations. For each case we have integrated the Langevin equation of motion of the system by using a stochastic Runge-Kutta algorithm \cite{runge-kutta} with $dt = 0.01$, which is small enough for all the calculations here performed.

The initial spatial configuration of the polymer is put with all the beads lying at the
rest distance $d_0=1$ with respect to each other, and the last sited
at coordinate $x_N(0)=0$, the final position of the external
force region. Every simulation stops when the coordinate of the center
of mass $x_{cm}=1/N \sum_i^{N} x_i$ of the chain reaches
the position $x_{cm}=0$ and the mean first passage time (MFPT) is then
computed as
 \be MFPT = \tau = \frac{1}{2\pi} \int_0^{2\pi}{\langle
  \tau(\phi) \rangle d \phi},
  \ee
where $\tau(\phi)$ represent the FPT of a single realization, and
$\phi$ is randomly chosen.

 \begin{figure}[htbp]
\centering
\includegraphics[angle=-90, width=8.5cm]{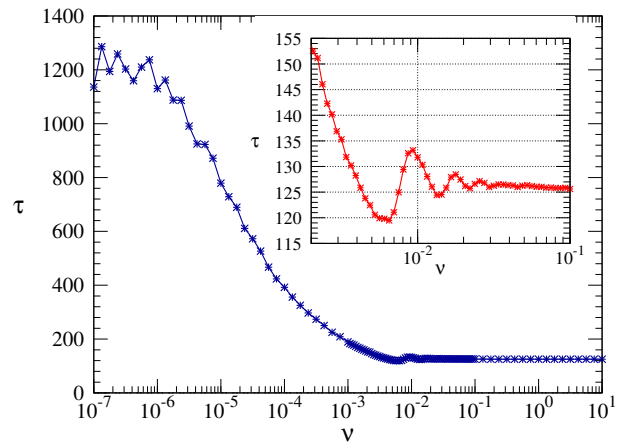}
\caption{(Color online) Log plot of the mean first passage time $\tau$
as a function of frequency of the oscillating force for the damping
 parameter $\gamma = 1$. The minimum and the oscillating behavior is
  showed in the inset. The thermal noise intensity is $D=0.01$. The
  first minimum, which is also the global one, satisfies the very
  robust condition: $\nu_m^{-1} = T_m = \alpha \tau_m$, were $\alpha$
  is a constant.} \label{RA}
 \end{figure}
The adoption of the center of mass as reference point in the movement has the role of avoiding the boundary effects given by the spatial extension of the motor. In fact at the end of the translocation process of the full chain, the total force acting on the polymer is smaller because of the decreasing number of monomers inside the motor. The choice of the center of mass as a reference will preserve the calculations from this unwanted boundary effect, being the average number of monomers inside the motor always the highest possible.
Fig.~\ref{RA} shows the MFPT $\tau$ as a function of the frequency of the oscillating force for damping parameter $\gamma = 1$, which
corresponds to a quite damped condition for the system. A clear
minimum is visible at a resonant frequency $\nu_m \sim 6\cdot
10^{-3}$. Moreover, a very strong oscillating behavior appears in the region $\nu \in [4\cdot 10^{-3} - 4\cdot 10^{-2}]$, before the high
frequency saturation value is reached.
 \begin{figure}[tbp]
\centering
\includegraphics[angle=-90, width=8.2cm]{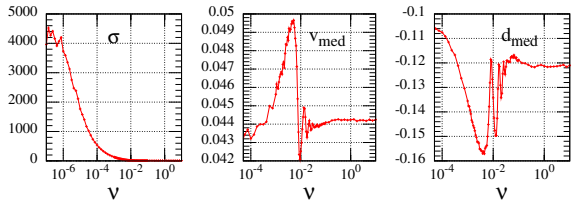}
\caption{(Color online) Standard deviation $\sigma$ (left), mean
  velocity (center), and mean difference between the rest distance
  $d_0$ and the distance of the monomers during the dynamics (right)
  related to Fig.\ref{RA}. The negative values of $d_{med}$ means that
  the springs are compressed. The oscillation appears in both the mean
  velocity and the elongation difference.} \label{RA2}
 \end{figure}
The thermal noise intensity is given by $D=0.01$.  Temperature
acts here in destructive way, hiding the oscillating information
contained into the MFPT output, as we will study later on (see in Fig.~\ref{D}).

Fig.~\ref{RA2} shows the standard deviation $\sigma$ (left panel), the
mean velocity (center), and the mean elongation $d_{med}$ of the
monomers with respect the rest distance $d_0$ during the dynamics
(right).  $d_{med}$ is calculated as the mean along the chain of the
distance between subsequent monomers, then further averaged in the
number experiments. The negative values of $d_{med}$ means that the
springs are in average compressed. The oscillation appears both in the
mean velocity and in the elongation. The minimum passage time is
obtained for the maximum velocity which also corresponds to the
minimum elongation of the springs, i.e. the maximum average
compression.

The computed value of $\tau$ has been averaged over the initial phase
$\phi$. In the low frequency limit we can compute the MFPT value as
\be
\tau_M =\frac{1}{2\pi} \int_0^{2\pi}{\tau_s(\phi) d \phi}.
\ee
\noindent Here $\tau_s$ is the value of the MFPT in the case of fixed
potential, being the low frequency limit an asymptotic limit where the
time scale of the dynamics is much faster than the period $T=1/\nu$ of
the external force. In that case, the polymer "sees" a static
potential in all his translocation time. Because the sliding is very
slow when the force is close to zero ($\phi= 0$, free diffusion
case), the MFPT tends to grow up and this explains its high values.

\begin{figure}[htbp]
\centering
\includegraphics[angle=-90, width=8.2cm]{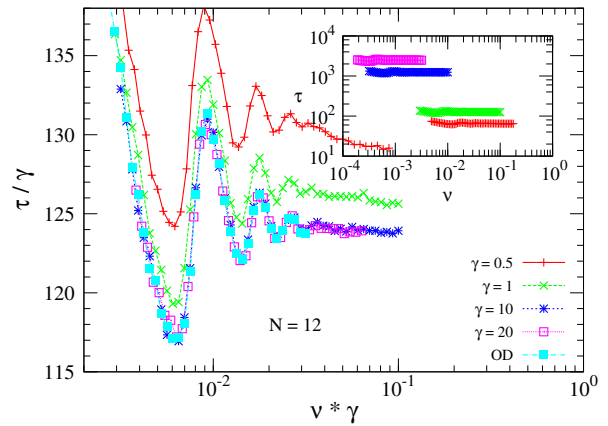}
\caption{(Color online) The minimum and the oscillating behavior is
  present independently on the damping parameter $\gamma$. The figure
  show the calculations for different $\gamma$ values: 0.5, 1, 10, 20,
  plotted following the scaling indicated in the legend of the
  axis. $\gamma=10$ approaches very well the overdamped limit (also
  shown). In the inset we show the curves without scaling.}
\label{gamma}
\end{figure}

In opposite way, the high frequency limit corresponds to a
dynamical translocation scale much slower than the period $T$. In
that case the polymer sees a constant mean force $F_0$, being the
cosine contribution averaged to zero during the slow dynamics.
In this case the MFPT can be easily evaluated
analytically:
 \be \label{vcm}
  v_{cm}(t) = \frac{n_{\gamma}F_0}{N \gamma} \left(1- e^{-\gamma t} \right)
 \ee
where $n_{\gamma}$ is the mean number of monomers at a given $\gamma$
inside the force region at the high frequency limit. Integrating
Eq.~\ref{vcm} in time we get:
 \be \label{Lcm}
  L_{cm}(\tau) = \frac{n_{\gamma} F_0}{N \gamma} \left[\tau + \frac{1}{\gamma}\left( e^{-\gamma \tau} -1 \right) \right]
 \ee
which is the distance covered by the polymer center of mass in the
time $\tau$. Unfortunately the equation is transcendent, and the
number $n_{\gamma}$ cannot be guessed in advance. In the overdamped
case ($\gamma \rightarrow \infty$) with also the related time
rescaling ($\tau' = \tau / \gamma $), the equation becomes:
 \be \label{LcmOD}
  L_{cm}(\tau') = \frac{n_{\gamma} F_0}{N} \tau'
 \ee
A rough estimation of $\tau'$ can be obtained using the rest value for
the number of monomers into the field, $n_{\gamma}=5.5$ and $L_{cm}$
as the distance between the initial position of the polymer center of
mass ($x_{cm}(0)=-L_{cm}$) and the position $x=0$ of the
boundary. Remembering the initial condition, $L_{cm} = (N-1)/2 = 5.5$,
we have, finally:
 \be \label{LcmOD}
  \tau' = \frac{N (N-1)}{2 n_{\gamma} F_0} \simeq 120,
 \ee
which is quite close to the high frequency value of Fig.~\ref{gamma},
where the overdamped limit is plotted together with other curves.

The intermediate region is characterized by a translocation time of
the same order of magnitude than the period of the external force,
$\tau \sim T$. In this special case the polymer can, in average, reach
the boundary during the first period of the driving. Thus the
interplay between the sliding down time and the external force period
reaches its maximum synchronization at this frequency
$\nu_m$. The second minimum is reached when the polymer escapes during
the $2^{\rm nd}$ period of the driving and so on. This explains why
the effect is a classical and not noise-induced: the synchronization
with the force driving occurs when the boundary is reached at the
proper time, and this can also occur without any fluctuations, though
the latter are unavoidable in the dynamics.

 \begin{figure}[tbp]
\centering
\includegraphics[angle=-90, width=8.2cm]{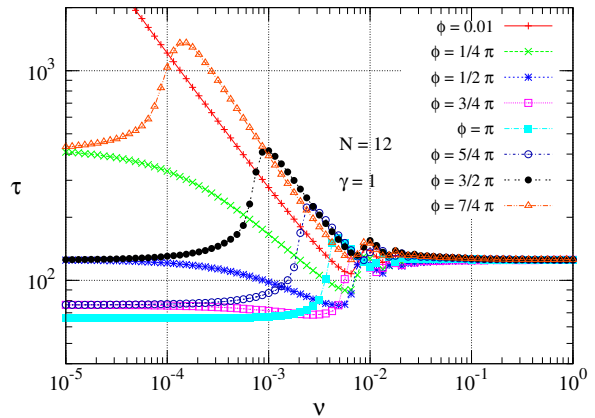}
\caption{(Color online) Mean first passage time $\tau$ as a function
  of frequency of the oscillating force for the damping parameter
  $\gamma = 1$ with various values of the initial phase $\phi$, namely
  $\phi=0.01, \pi/4, \pi/2, 3\pi/4, 5\pi/4, 3\pi/2, 7\pi/4 $. A
  minimum is always present. Moreover a very strong oscillating
  behavior appears in the region before the high frequency saturation
  value. The thermal noise intensity is $D=0.01$.} \label{phase}
 \end{figure}

For a deeper understanding of the basic dynamics, a set of simulations
have been done by fixing the initial phase, reported in
Fig.~\ref{phase}. We can see that the MFPT can show as well a
nonmonotonic behavior, depending on the $\phi$ value. According with
the previous comments, for low frequency values, we observe very high
$\tau$ values for $\phi=0.01$ (very small force during all the
dynamics), while for high frequency values again the static mean force
$F_0$ contributes to the dynamic, giving the same $\tau$ value for all
the different initial phases, which is also the value of the randomly
chosen initial conditions. The intermediate region shows first minima
or first maxima at different positions, depending on increasing or
decreasing initial forces respectively, followed again by
oscillations. It is worth to note that the couple of opposite initial
phases ($\phi=\pi/4$ and $\phi=5 \pi/4$, or $\phi=3\pi/4 $ and
$\phi=7\pi/4 $) show the same $\tau$ value in the low frequency limit,
because the force is the same nevertheless the direction (upward or
downward) of the cosine function is, and for the couple $\phi=\pi/2 $
and $\phi=3\pi/2 $ this limit is the same of the high frequency one,
because the limit force is $F_0$ in both cases.

\begin{figure}[tbp]
\includegraphics[angle=-90, width=8.2cm]{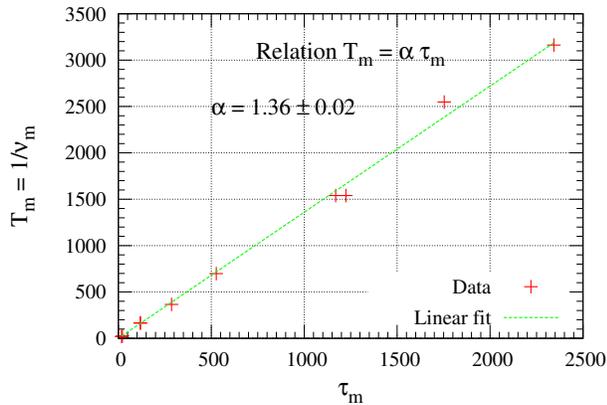}
\centering \caption{(Color online) Resonant MFPT ($\tau_m$) vs
  resonant period ($T_m$) for all the case investigated. The data
  concerns the parameters $\gamma = 0.01, 0.1, 1, 10, 15, 20$ and also
  the overdamped case with $N=12$. Furthermore, for $\gamma=1$, the
  different polymer length $N=12, 18, 24, 36$. All the points lie on
  the linear relation given by Eq.~\ref{eq.alpha}.} \label{alpha}
\end{figure}

As we mentioned previously, $\tau_m$, the first minimum of the MFPT
versus frequency curve, which is also the global minimum, satisfies
a very robust relation with the period of the driving for this
minimum, $T_m$, given by
\be
\label{eq.alpha}
1/\nu_m = T_m = \alpha \tau_m,
\ee
were $\alpha = 1.36 \pm 0.02$ (see Fig.~\ref{alpha}). This means that
the minimum translocation time is obtained for a driving period which
approximately corresponds to one and one third the time the polymer
takes to cross the boundary. The above condition has been checked for
all the values of $\gamma$ and all the polymer length we have studied.

As a further confirmation of Eq.~\ref{eq.alpha}, we show in
Fig.~\ref{gamma} the oscillating MFPT region for different values of
damping. In order to compare the curves, the MFPT and the frequency
have been scaled by damping. The presence of a minimum followed by a
series of oscillations is observed for all the values of the damping.
In fact, by diminishing the damping we obtain higher $\tau/\gamma$
ratios but the position of the minima and the amplitude of the
oscillations do not change. It shows that the robustness of the
oscillating pattern goes beyond Eq.~\ref{eq.alpha}.

Further, the presence of the fluctuations with different damping at
the same amplitudes (in the scaled values), demonstrate once more that
the oscillating behavior has nothing to do with inertial features and
inner vibrational chain modes, as one can infer at a first look of the
affect.
\subsection{Length scaling}\label{scaling}
Based on the hypothesis that in a polymer the friction is
proportional to $L^{\mu}$, for the driving case it has been
proposed \cite{sung96} that the translocation time of a
polymer follows a scaling law with respect to its length $L$ as $\tau
\propto L^{1+\mu}$, where $\mu$ is a certain constant. This doubtful
dependence has been recently discussed in Ref.~\cite{luo,chuang}.
\begin{figure}[bp]
\centering
\includegraphics[angle=-90, width=8.2cm]{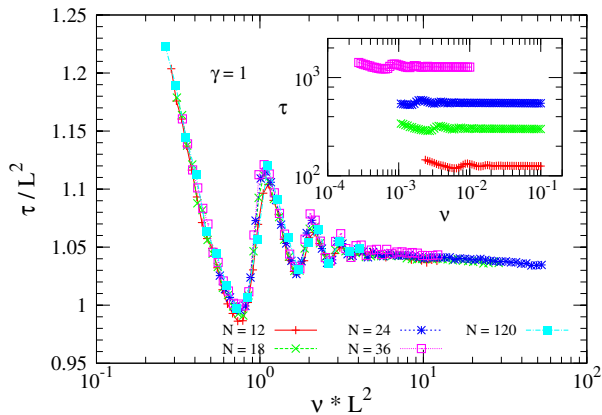}
\caption{(Color online) MFPT as a function of $\nu$ for different
  values of the polymer length $L$, namely $N = 12, 18, 24, 36,$ and
  120. When axis are properly scaled all the original curves (shown in
  the inset) superimpose upon each other.}
\label{N}
\end{figure}
In our unidimensional case the scaling law we recover appears to be in
good agreement with the Rouse prediction \cite{Rouse,deGennes}
\be
\tau \propto L^2.
\label {scaling}
\ee
Fig.~\ref{N} shows that the scaling properties hold here in a precise
way. There we show our numeric results for four different
polymer size, $N = 12, 18, 24, 36, 120$. It is observed that the
translocation times follow perfectly the scaling law $\tau \propto
(N-1)^2$ where $N-1$ is proportional to the total length: $L=d_0(N-1)$
with $d_0$ the rest distance between two subsequent
monomers. This relation confirms again the robustness of
Eq.~\ref{eq.alpha}. The scaling law given in Eq.~\ref{scaling} has been also proved by other works where a constant force was used to drive the polymer translocation in a 1d case \cite{chuang,Grosberg}.

\subsection{Thermal noise contribution}\label{noise}
\begin{figure}[htbp]
\centering
\includegraphics[angle=-90, width=8.2cm]{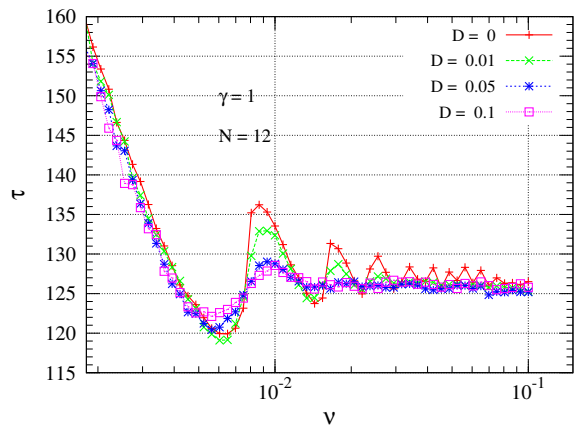}
\caption{(Color online) Range of the oscillating effect for
different thermal noise intensity, namely: $D = 0, 0.1,0.5, 0.1$.
The oscillating behavior tends to disappear by increasing the
temperature. The maxima decrease and the minima increase.}
\label{D}
\end{figure}
The oscillating behavior depends on the thermal noise in a destructive
way. As we can see in Fig.~\ref{D}, the role of the thermal noise then
appears to be to destroy the oscillations. These are present in the
classical dynamics (no noise), but the number of visible maxima
decreases by increasing the noise intensity, and the value of its
maxima decreases monotonically. Only the first maximum remains barely
visible in the plot at the highest temperature values there shown.  It
has to be observed that the behavior shown in Fig.~\ref{D} for high
values of $D$, is qualitatively the same of the typical resonant
activation (RA) phenomenon \cite{RA}, where a particle overcomes a
potential barrier. RA effect has been observed also in polymer
dynamics \cite{Pizz2} with a characteristic curve very similar to the
behavior shown here.  In this sense an experimental curve showing
similar behavior as a function of the frequency to the one here
described, can be related to the presence of a fluctuating barrier or,
as in the case here presented, to an oscillating linear driving in an
high temperature environment.

\subsection{Stall Force}\label{stallforce}

\begin{figure}[htbp]
\centering
\includegraphics[width=8.5cm]{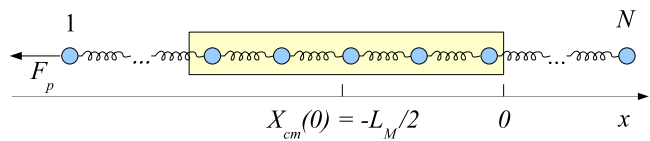}
\caption{(Color online) A force pull is applied at the first monomer
  to measure the stall force of the potential.} \label{stall2}
\end{figure}
A last result of the model here studied, useful for future analysis
and comparisons with other models or experimental outcomes, is the
evaluation of the stall force, i.e. the force necessary to stop the
polymer translocation. In order to do that, a set of simulation have
been performed by applying a pulling force $F_p$ (See
Fig.~\ref{stall2}) on the left side of the chain. The initial
condition has been fixed now with the polymer center of mass in the
center of the external potential, and the velocity of the center of
mass is measured waiting for the exit on the left or on the right of
the potential region. A zero mean velocity allows to estimate the
stall force of the system.

As it is visible in the inset of Fig.~\ref{RA}, the mean velocity of
the polymer shows a strong oscillating behavior, similar for all the
$\gamma$ values here used.

\begin{figure}[tbp]
\centering
\includegraphics[angle=-90, width=8.2cm]{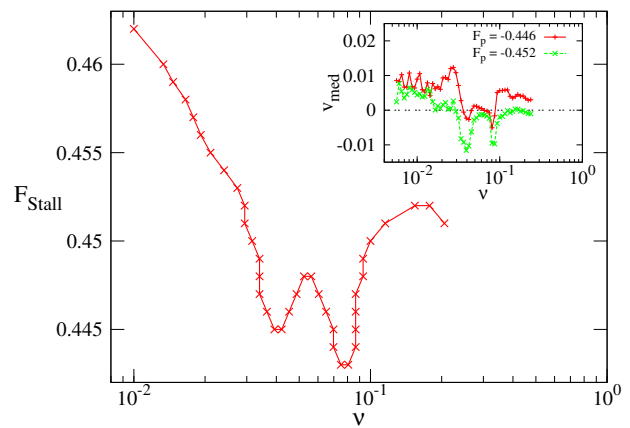}
\caption{(Color online) Stall force as a function of the frequency of the oscillating driving for $\gamma = 0.1$. The other parameters are
the same of Fig.~\ref{RA}. The inset shows the mean velocity versus
frequency curves for two values of the pull force, $F_p=-0.446$ and $-0.452$.}
\label{fv-g01}
\end{figure}

\begin{figure}[htbp]
\centering
\includegraphics[angle=-90, width=8.2cm]{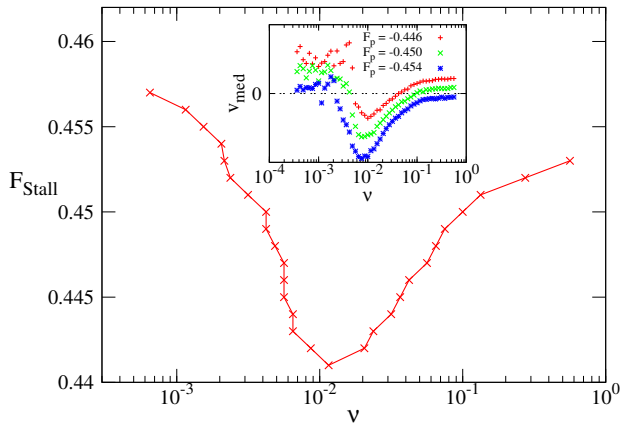}
\caption{(Color online) Stall force as a function of the frequency of  the oscillating driving for $\gamma = 1$. The two minima displayed  in Fig.~\ref{fv-g01} have disappeared, and now only a minimum  appears. The other parameters are the same of Fig.~\ref{RA}. The  inset shows the mean velocity versus frequency curves for three  values of the pull force, $F_p=-0.446$, $-0.450$, and $-0.454$.}
\label{fv-g1}
\end{figure}

Surprisingly, the oscillations are not always present in the velocity used for the stall force evaluation. In fact for $\gamma = 0.1$ they
are present, while for $\gamma = 1$ they are not. The inset of the two figures \ref{fv-g01} and \ref{fv-g1} show the two cases. Similarly the two related stall forces present different behavior. With two minima the one with $\gamma =0.1$ (Fig.~\ref{fv-g01}), with only one the other with $\gamma = 1$ (Fig.~\ref{fv-g1}).
In both cases a strong mean frequency dependence of the stall force is depicted, though the intensity of force variation appears relatively
not very high. In fact the variation in the scale is somehow small,
and an experimental verification could be not immediately simple to
perform. In this sense the more realistic model depicted in
\cite{fjf-damn} is much more checkable.

\section{Conclusions}
We have studied the main characteristics of a simple toy model for
studying the translocation time of a polymer driven by a uniform force
oscillating sinusoidally in time and acting on a limited spatial
region. The translocation time, as well as the stall force of the
system show a strong dependence on the frequency of the driving. This
dependence is visually similar, in many respects, and with many
\textit{distinguo}, to the resonant activation phenomenon.  The
driving frequency for the minimum translocation time follows with
strict proportionality the computed value of this very time. A simple
scaling law for the polymer length, $\tau \propto L^2$ has been found.
The role of the noise intensity has been also investigated, which
gives, in this case, destructive effects on the registered MFPT
oscillating behavior. Finally, we have studied the stall force of the
motor as a function of the driving frequency, finding a clear non
monotonic behavior.

We present here results for a one-dimensional model. Further investigations will consider dimensionality effects on the quantities here calculated. Dimensionality could be irrelevant for polymers constrained to move in confined channels. Moreover, in many experimental situation the polymer is stretched, thus removing the dimensionality dependence of the measured quantities.

 The model can have artificial application in
nanotechnological devices driven by oscillating fields, and represents a first step to further analysis and applications to
realistic biological translocation processes.

\vspace{0.5cm}
The authors acknowledge the financial support from Spanish MICINN through Project No. FIS2008-01240, cofinanced by FEDER funds.



\begin{thebibliography}{99}

\bibitem{KasPNAS96}
J. J. Kasianowicz, E. Brandin, D. Branton, and D. W. Deamer, {\it
Proc. Natl. Acad. Sci. USA} {\bf 93}, 13770 (1996)

\bibitem{Pizz}
N. Pizzolato, A. Fiasconaro, B. Spagnolo, {\it Int. J. Bifurc.
Chaos} {\bf 18}, 2871 (2008); N. Pizzolato, A. Fiasconaro, and B. Spagnolo, {\it J. Stat. Mech: Theory and Exp.} P01011 (2009).

\bibitem{linke}
M.T. Downton, M.J.Zuckermann, E.M.Craig, M.Plischke, H.Linke, {\it Phys. Rev. E} {\bf 73}, 011909 (2006);
E.M.Craig, M.J.Zuckermann, H.Linke, {\it Phys. Rev. E} {\bf 73},
051106 (2006)

\bibitem{Bust}
D. E. Smith, S. J. Tans, S. B. Smith,
S. Grimes, D. L. Anderson, and  C. Bustamante, {\it
Nature} {\bf 413}, 748 (2001)

\bibitem{gut}
B. Guti\'errez-Medina, A. N. Fehr, and  S.M. Block, {\it
Proc. Natl. Acad. Sci.} {\bf 413}, 748 (2001)

\bibitem{mickler}
M. Mickler, E. Schleiff, and T. Hugel {\it Chem. Phys. Chem.} {\bf 9}, 1503 (2008).

\bibitem{RA}
C.R. Doering, and J. C. Gadoua, {\it Phys. Rev. Lett.} {\bf 69},
2318 (1992);
M. Bier, and R.D. Astumian, {\it Phys. Rev. Lett.} {\bf 71}, 1649
(1993);
M. Bogu\~n\'a, J. M. Porra, J. Masoliver, and K. Lindenberg, {\it Phys. Rev. E} {\bf 57}, 3990 (1998);
R. N. Mantegna and B. Spagnolox, {\it Phys. Rev. Lett.} {\bf 84},
3025 (2000); A. A. Dubkov, N. V. Agudov, and B. Spagnolo, {\it Phys. Rev. E} {\bf 69}, 061103 (2004).

\bibitem{Rouse}
P. E. J. Rouse, {\it J. Chem. Phys.} {\bf 21}, 1272 (1953).

\bibitem{sung96}
W. Sung and P. J. Park,  {\it Phys. Rev. Lett. }{\bf77}, 783
(1996).

\bibitem{chuang}
J. Chuang, Y. Kantor, and M. Kardar {\it Phys. Rev. E }{\bf 65}, 011802 (2001).

\bibitem{luo}
K. Luo, T. Ala-Nissila, S.C. Ying, and R. Metzler,  {\it Eur. Phys. Lett.}{\bf 88}, 68006 (2009).

\bibitem{deGennes}
P-G de Gennes, Scaling Concepts in Polymer Physics (Cornell University Press, Ithaca and London, 1979).

\bibitem{Grosberg}
A.Yu. Grosberg, S.Nechaev, M. Tamm, and O. Vasilyev, {\it Phys. Rev. Lett.} {\bf 96}, 228105 (2006)

\bibitem{Pizz2}
N. Pizzolato, A. Fiasconaro, D. Persano Adorno, B. Spagnolo, {\it Phys. Biol.}{\bf 7} 034001 (2010).

\bibitem{runge-kutta}
H. S. Greenside and E. Helfand, {\it Bell Sistem Technical Journal}, {\bf 60} 1927 (1981).

\bibitem{fjf-damn}
A. Fiasconaro, F. Falo, and J.J.Mazo, {\it unpublished}.

\end{thebibliography}
\end{document}